\documentclass[runningheads]{llncs}
\usepackage[T1]{fontenc}
\usepackage{graphicx}
\usepackage{booktabs}
\usepackage[misc]{ifsym}
\newcommand{\corr}{(\Letter)}
\usepackage{algorithm}
\usepackage{algorithmicx}
\usepackage{algpseudocode}
\usepackage{multirow}
\usepackage{tabularx}
\usepackage[table]{xcolor}
\usepackage{amsmath}
\usepackage{amssymb}
\usepackage{makecell}
\usepackage{array}
\usepackage{hyperref}
\usepackage[para,online,flushleft]{threeparttable}

\definecolor{gain}{HTML}{FF7F00} % Cyan color for gains
\definecolor{loss}{HTML}{009B9E} % Orange color for losses

\begin{document}
\title{MOTOR: Learning ID-free Item Representation with Token Crossing for Embedding-based Multimodal Recommendation}

\author{Kangning Zhang\inst{1} \and
Jiarui Jin\inst{2} \and
Jianghao Lin\inst{1} \corr \and
Yingjie Qin\inst{2} \and
Weinan Zhang\inst{1} \corr \and
Yong Yu\inst{1} \corr
}
\authorrunning{K. Zhang et al.}
\institute{Shanghai Jiao Tong University, Shanghai, China\\
\email{\{zhangkangning,linjianghao,wnzhang\}@sjtu.edu.cn}\\
\email{yuyong@apex.sjtu.edu.cn}
\and 
Xiaohongshu Inc., Shanghai, China \\
\email{\{jinjiarui,huanling\}@xiaohongshu.com}
}
\titlerunning{MOTOR: Learning ID-free Item Representation}
\maketitle              % typeset the header of the contribution

\begin{abstract}
While multimodal recommendation models have effectively integrated visual and textual information, their reliance on unique ID embeddings constitutes a fundamental performance bottleneck. Specifically, ID-based paradigms suffer from three limitations: (1) \textbf{Information Isolation}, where unique IDs prevent semantic information exchange among related items; (2) \textbf{Cold-Start Vulnerability}, as ID embeddings are difficult to optimize with sparse interactions; and (3) \textbf{Storage Inefficiency}, where parameter costs scale linearly with item quantity. To overcome these challenges, we propose \textbf{MOTOR}, a novel \textbf{ID-free MultimOdal TOken Representation} scheme. MOTOR replaces explicit item IDs with learnable, shared multimodal tokens, fundamentally transforming the recommender into an ID-free framework. Methodologically, we first employ product quantization to discretize raw multimodal features into compact token IDs. These tokens serve as implicit item features, which are then synthesized via a novel \textbf{Token Cross Network (TCN)} to capture high-order interaction patterns. This "discretize-and-interact" mechanism enables semantic sharing across items and significantly compresses the model size without introducing complex auxiliary losses. Extensive experiments
across nine mainstream models demonstrate the significant performance
improvement achieved by MOTOR. Further, MOTOR improves the capability of these models to recommend items in cold-start scenarios.
\keywords{Multimodal Recommendation \and ID-Free Recommendation\and Tokenization}
\end{abstract}
\section{Introduction}\label{sec:intro}

Recommender systems serve as the fundamental engine for content discovery in the era of information overload. While traditional collaborative filtering (CF) struggles with data sparsity, Multimodal Recommender Systems (MRSs) have emerged as a robust solution by incorporating item-related multimedia features (e.g., images, text) to enrich item representations~\cite{vbpr,dualgnn,BM3,Freedom,MGCN,AlignRec,DREAM,Dual-Aligned}.
Existing approaches typically fuse these multimodal signals with unique item ID embeddings via concatenation, summation~\cite{vbpr,liu2017deepstyle}, or Graph Neural Networks (GNNs)~\cite{wei2019mmgcn,dualgnn,GRCN,MGCN,AutoGraph} to capture high-order semantics. Despite their success, these models predominantly adhere to an \textbf{ID-centric paradigm}, where multimodal features merely serve as auxiliary information to assist the learning of unique ID embeddings.

However, the heavy reliance on ID embeddings introduces intrinsic limitations that hinder the scalability and robustness of current recommenders:
(i) \textbf{Information Isolation}: ID embeddings are inherently orthogonal and independent. This prevents the direct propagation of collaborative signals among semantically related items, forcing the model to rely solely on interaction overlap to learn similarities;
(ii) \textbf{Cold-Start Vulnerability}: ID embeddings are essentially look-up parameters that require sufficient interaction data to be optimized. For cold-start or long-tail items, these embeddings remain under-trained, creating a performance bottleneck that multimodal features alone cannot fully compensate for;
(iii) \textbf{Storage Inefficiency}: The parameter space for ID embeddings scales linearly with the item inventory ($O(N)$), imposing significant memory burdens as the system grows.

\begin{figure*}[ht]
  \centering
  \includegraphics[width=1.0\textwidth]{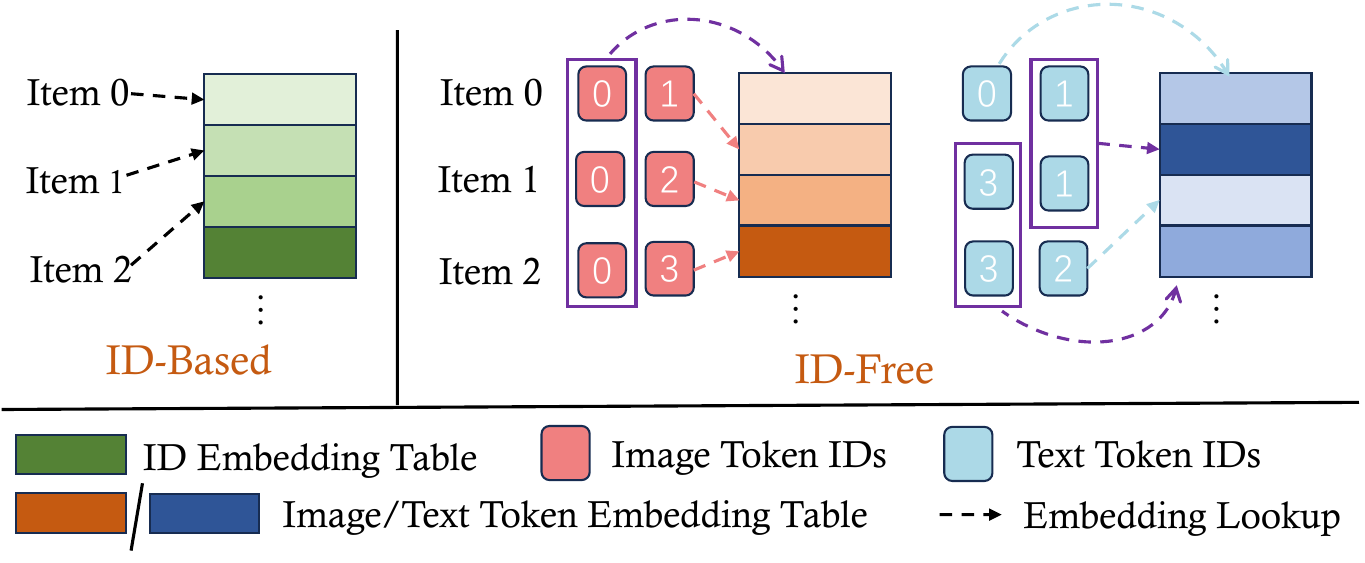}
  \caption{
MOTOR transforms the ID-centric recommendation model into an ID-free framework by removing the ID Embedding Table and maintaining modal-specific Token Embedding Tables with fewer parameters. Consider a scenario where items 0 and 1 are popular, while item 2 is a cold-start item. MOTOR establishes connections among these items through shared tokens in either the image or text modality. Consequently, cold-start item 2 can enhance its representation through sharing tokens with other related items.
  }
  \label{fig:ID_basedvsID_free}
\end{figure*} 

To address these challenges, we propose a paradigm shift from ID-centric to ID-free recommendation. We introduce \textbf{MOTOR}, a streamlined framework that replaces unique ID embeddings with learnable, shared multimodal token embeddings, as illustrated in Figure \ref{fig:ID_basedvsID_free}.
The core intuition is that items should be represented by a composition of shared semantic units rather than isolated identifiers. Specifically, MOTOR employs optimized product quantization (OPQ)~\cite{OPQ} to discretize continuous multimodal features into a finite set of shared token IDs.
This design fundamentally resolves the aforementioned limitations:
First, semantically similar items naturally share a subset of tokens, breaking \textbf{Information Isolation} and enabling knowledge transfer.
Second, for the \textbf{Cold-Start} problem, new items can immediately ``borrow" well-trained token representations from popular items with similar content, bypassing the need for extensive interaction history.
Third, by maintaining compact Token Embedding Tables instead of a massive ID table, MOTOR significantly alleviates the \textbf{Storage Burden}. Empirically, MOTOR requires only $\sim$20\% of the original parameters to achieve superior performance on large-scale datasets.

Merely replacing IDs with tokens is insufficient; capturing the complex dependencies between these discrete units is crucial. In MOTOR, we treat token embeddings as \textit{implicit item features} and design a lightweight \textbf{Token Cross Network (TCN)} to model their interactions. Unlike complex graph structures, TCN efficiently computes first-order, second-order, and high-order feature interactions to generate a comprehensive Token Representation.
We further explore two interaction strategies: \textit{Modal-specific} (interacting within modalities) and \textit{Modal-agnostic} (interacting across all modalities globally).
Crucially, MOTOR is designed as a plug-and-play module: the learned Token Representations can directly replace ID embeddings in existing multimodal recommenders without requiring additional loss functions or complex architectural changes.

Our main contributions are summarized as follows:
\begin{itemize}
    \item We identify the critical bottlenecks of the ID-centric paradigm and propose MOTOR, a novel scheme that pioneers the integration of tokenization mechanisms into embedding-based multimodal recommendation, establishing a shift toward ID-free systems.
    \item We propose a ``Tokenize-then-Cross'' architecture. By treating tokens as implicit features and employing a lightweight Token Cross Network (TCN), we effectively capture multi-order interactions within and across modalities. We systematically evaluate both Modal-specific and Modal-agnostic interaction strategies.
    \item We conduct extensive experiments across five datasets and nine backbone models in both Common and Cold-Start scenarios. Results demonstrate that MOTOR consistently yields significant performance gains, particularly for long-tail items, while drastically reducing storage costs.
\end{itemize}

\section{Related Work}\label{sec:related}

\textbf{ID-based Multimodal Recommendation.}\par\noindent
ID-based multimodal recommenders combine item ID embeddings with visual and textual signals to model user preferences. Representative methods include VBPR~\cite{vbpr}, graph-based models such as MMGCN and DualGNN~\cite{wei2019mmgcn,dualgnn}, and self-supervised alignment approaches such as BM3 and SLMRec~\cite{BM3,SLMRec}. Other multi-view or graph-enhanced designs include FREEDOM, MGCN, AlignRec, DREAM, DAMORE, and AutoGraph~\cite{Freedom,MGCN,AlignRec,DREAM,Dual-Aligned,AutoGraph}. Nevertheless, these methods still rely on high-quality item ID embeddings. Poor ID optimization can markedly impair performance even with multimodal cues.

\textbf{ID-free Recommendation.}\par\noindent
ID-free recommendation learns inductive user/item representations from content or interaction structure, avoiding explicit ID embeddings. Prior work includes relation estimation from interactions~\cite{wu2021towards}, template-based inductive embeddings such as INMO~\cite{INMO}, sequence encoders over item text such as Recformer~\cite{Recformer} and UniSRec~\cite{UniSRec}, and pre-trained unimodal item encoders such as IDvsMoRec~\cite{IDvsMoRec}. ID-free multimodal recommendation remains insufficiently studied. MOTOR fills this gap by using learnable multimodal token representations to replace item ID embeddings, thereby transforming standard ID-based multimodal recommenders into ID-free systems.

\section{Methodology}
% \begin{figure}[h]
% % \setlength{\abovecaptionskip}{-0cm}
% % \setlength{\belowcaptionskip}{-0.5cm}
%   \centering
%   \includegraphics[width=0.50\textwidth]{figs/MOTOR.pdf}
%   \caption{
%   A high-level illustrate of MOTOR. (a) The framework of MOTOR, which includes: Product Quatization (PQ), Token Embeddings, Token Cross Network; (b) Any Multimodal Recommendation Models as backbone models to integrate the MOTOR and complete the recommendation task; (c) The Token Cross Network is illustrated in detail in Figure \ref{fig:Token Cross Network} and Section \ref{sec:TCN}. 
%   }
%   \label{fig:MOTOR}
% % \vspace{-5pt}
% \end{figure}

\begin{figure*}[ht]
  \centering
  \includegraphics[width=1.0\textwidth]{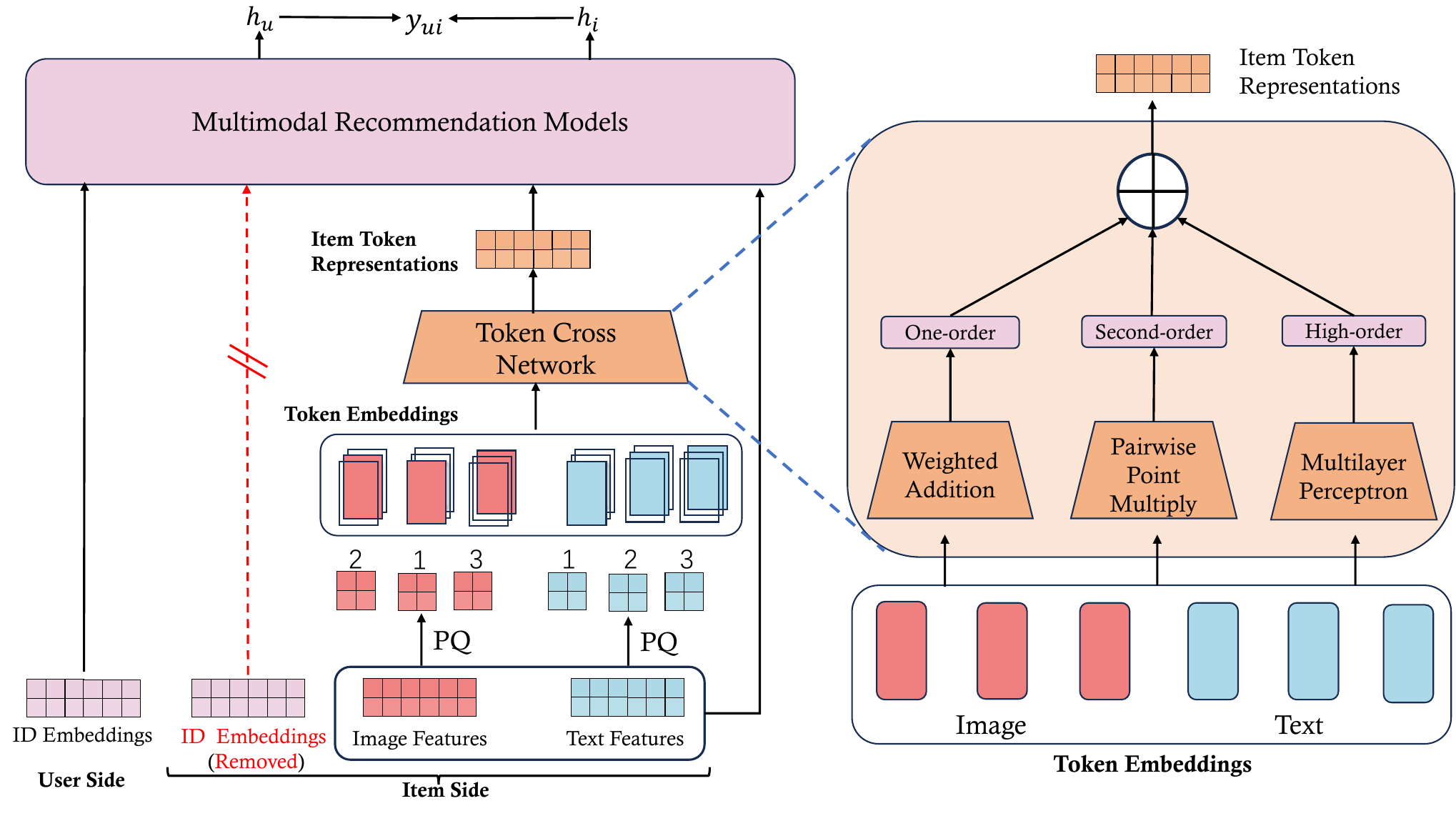}
  \caption[A high-level illustration of MOTOR.]{
  A high-level illustration of MOTOR. 
The dashed line over the Item ID Embedding indicates that MOTOR replaces the original ID embeddings with the learned Item Token Representations. The token embeddings and Token Cross Network are optimized along with the downstream recommenders in an end-to-end manner. The core components of MOTOR include Feature Discretization (Section \ref{sec:Token Learning}), Token Embeddings (Section \ref{sec:token embeddings}), and Token Cross Network (Section \ref{sec:Token Cross Network}). In the diagram, the Token Cross Network on the right is Modal-agnostic, which performs a holistic cross-fusion of tokens from all modalities.}
  \label{fig:MOTOR}
% \vspace{-5pt}
\end{figure*}

% \begin{figure}[h]
% % \setlength{\abovecaptionskip}{-0cm}
% % \setlength{\belowcaptionskip}{-0.5cm}
%   \centering
%   \includegraphics[width=0.50\textwidth]{figs/TCN_v1.pdf}
%   \caption{
%   A high-level illustrate of MOTOR. (a) The framework of MOTOR, which includes: Product Quatization (PQ), Token Embeddings, Token Cross Network; (b) Any Multimodal Recommendation Models as backbone models to integrate the MOTOR and complete the recommendation task; (c) The Token Cross Network is illustrated in detail in Figure \ref{} and Section \ref{}.
%   }
%   \label{fig:MOTOR}
% % \vspace{-5pt}
% \end{figure}

\subsection{Preliminary}
Consider a set of $M$ users denoted by $\mathcal{U} = \{ u_j \}_{j=1}^{M}$, and a set of $N$ items denoted by $\mathcal{I} = \{ i_t \}_{t=1}^{N}$. Each item is associated with multimodal features $f_i^m \in \mathbb{R}^{d_m}$, where $m \in \{v, t\}$. In this context, $v$ and $t$ correspond to the visual and textual modalities, respectively.

The interaction records is represented as $R \in \mathbb{R}^{M \times N}$, where $R_{ui} = 1$ signifies that user $u$ has interacted with item $i$, and $R_{ui} = 0$ otherwise. 
% Essentially, $\mathcal{R}$ can be viewed as a sparse behavior graph $\mathcal{G} = (\mathcal{V}, \mathcal{E})$, with $\mathcal{V} = \mathcal{U} \cup \mathcal{I}$ representing the vertices, and $\mathcal{E} = \{ (u, i) \mid u \in \mathcal{U}, i \in \mathcal{I}, \mathcal{R}_{ui} = 1\}$ representing the edges. 
The set of items that a user $u$ has interacted with is denoted by $\mathcal{I}_u = \{ i \mid R_{ui} = 1 \}$. ID-based multimodal recommendation aims to precisely estimate user preferences by ranking items for each user, utilizing the predicted preference scores $y_{ui}$:
\begin{equation}
\begin{aligned}
    h_u, h_i &= f_{\theta}(e_u, e_i, f_i^t, f_i^v) \\
    y_{ui} &= (h_u)^Th_i
\end{aligned}
\end{equation}
In this function, $\theta$ represents the recommendation model's parameters, $e_u$ and $e_i$ stand for the ID embeddings, $h_u$ and $h_i$ mean the final user and item representations. In MOTOR, we endeavor to remove item ID embeddings and replace them with more lightweight token embeddings and TCN networks. The preference function of ID-free recommendation is denoted as:
\begin{equation}
\begin{aligned}
    r_i &= g_{\phi}(t_i^t, t_i^v) \\
    h_u, h_i &= f_{\theta}(e_u, r_i, f_i^t, f_i^v) \\
    y_{ui} &= (h_u)^Th_i
\end{aligned}
\end{equation}
where $r_i$ means the Token Representation output by TCN network, $\phi$ represents the parameters of token embeddings and TCN, and $t_i^m$ denotes the token IDs of item $i$ in modality $m$, with $m \in \{v, t\}$. \textbf{The parameters $\theta$ of the recommendation model, along with the parameters of the token embeddings and the TCN network $\phi$, are jointly trained in an end-to-end manner.} \footnote{We use lowercase letters to denote the vector corresponding to a specific item, and uppercase letters to represent the feature matrix of all items. The superscript denotes a specific modality.}
% For example, $r_i \in \mathbb{R}^{d}$ denotes the final Token Representation of item $i$, while $R \in \mathbb{R}^{N \times d}$ represents the final Token Representations of all items.} .

% In MOTOR, we attempt to learn a novel Item Token Representation $r_i$, to replace the item's ID embedding, $e_i$. The preference function of ID-free recommendation can be denoted as $y_{ui} = f_{\theta}(e_u, r_i, f_i^m)$\footnote{In this work, we use lowercase letters with subscripts to denote the vector corresponding to a specific item, and uppercase letters to represent the feature matrix of all items. For example, $r_i \in \mathbb{R}^{d}$ denotes the final Token Representation of item $i$, while $R \in \mathbb{R}^{N \times d}$ represents the final Token Representations of all items.}.

\subsection{The Framework of MOTOR}
The structure of MOTOR is illustrated in Figure \ref{fig:MOTOR}.
The core objective of MOTOR is to effectively learn $r_i$. This learning process can be divided into two steps. Firstly, by applying Product Quantization to the raw multimodal features of items, we obtain items' unique multimodal Token IDs $T^m$ in modality $m$. Based on the learned Token IDs, we can retrieve the token embeddings for specific items by looking them up in the Token Embedding Table. Here, each item corresponds to multiple token embeddings. Secondly, we treat the token embeddings obtained for each item as its implicit feature vectors. Through a lightweight Token Cross Network, we perform one-order, second-order, and high-order feature interactions on these implicit feature vectors, ultimately generating the Token Representation $r_i$. Next, we present the MOTOR in detail.

\subsection{Discretized Token Learning}\label{sec:Token Learning}
The initial textual and visual information of items can be encoded into specific feature vectors $F^t$ and $F^v$. Text features can be obtained with text encoders~\cite{bert,GPT2,sentencebert,T5}, and visual features can be obtained with visual encoders~\cite{VGG,resnet,vit,CLIP}. Within the MOTOR framework, we do not rely on specific encoders.

We consider mapping the modal features into discretized tokens. Taking the visual feature $F^v$ as an example, a similar approach is applied to the text feature $F^t$. 
We begin by partitioning visual feature $F^v \in \mathbb{R}^{N \times d_v}$ into $D$ subvectors, each of dimensionality $N \times \frac{d_v}{D}$. This can be expressed as:
\begin{equation}
    F^v = [F^v_1, F^v_2, \ldots, F^v_D]
\end{equation}
For each subvector $F_x^v$, we apply the Optimized Product Quantization (OPQ)~\cite{OPQ} to generate $K$ cluster centroids, constructing a codebook $C_x^v \in \mathbb{R}^{K \times \frac{d_v}{D}}$:
\begin{equation}
       C_x^v = \{ c_{x,1}^v, c_{x,2}^v, \ldots, c_{x,K}^v \}
\end{equation}
where $c_{x,j}^v \in \mathbb{R}^{\frac{d_v}{D}}$ represents the $j$-th centroid in the $x$-th subvector's codebook. 
% To acquire these PQ centroid embeddings, we employ a well-known technique called Optimized Product Quantization (OPQ)~\cite{OPQ}. 
% We optimize these centroid embeddings through a well-known technique called Optimized Product Quantization (OPQ)~\cite{OPQ}, utilizing the visual features of the items from the entire training dataset. 
Each subvector matrix $F^v_x$ is then quantized by finding the nearest centroid in the corresponding codebook $C_x^v$. The index of this centroid acts as the Token IDs. This assignment is expressed as:
\begin{equation}
\begin{aligned}
     t_{x,i}^v &= \text{argmin}_j \| f^{v}_{x,i} - c_{x,j}^v \| \\
     t_i^v &= [t_{1,i}^v, \ldots , t_{D,i}^v]
\end{aligned}
\end{equation}
where $t_{x,i}^v$ is the Token ID of item $i$ for the $x$-th subvector in vision modality. $t_i^v$ is the visual Token IDs for item $i$.
Finally, the discrete Token IDs $T^v \in \mathbb{R}^{N \times D}$ of all items are represented as a concatenation of $t_i^v$:
\begin{equation}
    T^v = [t_1^v, t_2^v, \ldots, t_N^v]
\end{equation}

In summary, through the process of vector clustering and quantization, visual features $F^v \in \mathbb{R}^{N \times d_v}$ is transformed into a set of discrete Token IDs $T^v \in \mathbb{R}^{N \times D}$. A similar procedure can be applied to convert text features $F^t$ into their Token IDs $T^t$. \textbf{The process of vector quantization requires execution only a single time and may subsequently be employed by all downstream models.}

\subsection{Token Embeddings as Implicit Features}\label{sec:token embeddings}
Given the learned discrete token IDs, we obtain token embeddings via embedding lookup.

For visual tokens $T^v \in \mathbb{R}^{N \times D}$, we construct $D$ embedding tables $\{E_x^v\}_{x=1}^{D}$, one per token-ID dimension. For each dimension $x$, all items share the same table $E_x^v \in \mathbb{R}^{K \times d}$, where $K$ is the number of cluster centers defined in Section~\ref{sec:Token Learning} (i.e., the index range), and $d$ is the embedding size. In MOTOR, $d$ matches the original ID-embedding dimension. For item $i$ with visual token IDs $t_i^v=[t_{1,i}^v,\ldots,t_{D,i}^v]$, the corresponding embeddings are $[e_{1,i}^v,\ldots,e_{D,i}^v]$, where $e_{x,i}^v$ is the $t_{x,i}^v$-th row of $E_x^v$. These token embeddings are distinct from PQ centroids. Similar to ID embeddings, we randomly initialize all token embedding tables and train them end-to-end with the downstream multimodal recommender. Text token embeddings are obtained analogously, yielding the overall token-embedding set for item $i$:
$\{e_{1,i}^v,\ldots,e_{D,i}^v, e_{1,i}^t,\ldots,e_{D,i}^t\}$.

In MOTOR, we treat these token embeddings as implicit features and model their interactions via the Token Cross Network. Intuitively, shared token IDs indicate shared latent semantic attributes, akin to belonging to the same category.

\subsection{Token Cross Network}\label{sec:Token Cross Network}
Feature interaction is central to CTR modeling~\cite{cheng2016widedeeplearning,DeepFM,DCN}. MOTOR treats item token embeddings as implicit features and employs a lightweight Token Cross Network (TCN), following DeepFM~\cite{DeepFM}, to capture first-, second-, and high-order interactions.

Because an item contains tokens from multiple modalities, we consider two interaction schemes: \textit{modal-specific} (MS) TCN, which crosses tokens within each modality and then aggregates modalities by summation, and \textit{modal-agnostic} (MA) TCN, which performs joint interactions over all tokens to directly produce the final token representation.

\subsubsection{\textbf{Modal-specific Token Cross Network}}
Using vision as an example, given visual token embeddings $\{e_{x,i}^v\}_{x=1}^{D}$, the first-order term is a weighted sum:
\begin{equation}
r^v_{\text{one}, i}=\sum_{x=1}^{D} w_x^v e_{x,i}^v .
\end{equation}
The second-order term reuses the same weights:
\begin{equation}
r^v_{\text{second}, i}=\sum_{x=1}^{D}\sum_{y=x+1}^{D} w_x^v w_y^v \left(e_{x,i}^v \odot e_{y,i}^v\right),
\end{equation}
where $\odot$ denotes element-wise multiplication. High-order interactions are modeled by an MLP over the concatenated embeddings:
\begin{equation}
\begin{aligned}
h_{0,i}^v &= \text{concat}(e_{1,i}^v,\ldots,e_{D,i}^v),\\
h_{l+1,i}^v &= \sigma(W_l^v h_{l,i}^v+b_l^v),\quad l=0,\ldots,L-1,\\
r^v_{\text{high}, i} &= h_{L,i}^v ,
\end{aligned}
\end{equation}
with activation $\sigma$ and depth $L$. The vision representation is
\begin{equation}
r_i^v=r^v_{\text{one}, i}+r^v_{\text{second}, i}+r^v_{\text{high}, i}.
\end{equation}
Text is processed analogously to obtain $r_i^t$, and the final item representation is
\begin{equation}
r_i=r_i^v+r_i^t .
\end{equation}

\subsubsection{\textbf{Modal-agnostic Token Cross Network}}
MA-TCN jointly crosses all token embeddings across modalities. For vision+text, there are $2D$ embeddings $\{e_{x,i}\}_{x=1}^{2D}$:
\begin{equation}
\begin{aligned}
r_{\text{one}, i} &= \sum_{x=1}^{2D} w_x e_{x,i},\\
r_{\text{second}, i} &= \sum_{x=1}^{2D}\sum_{y=x+1}^{2D} w_x w_y (e_{x,i}\odot e_{y,i}),\\
h_{0,i} &= \text{concat}(e_{1,i},\ldots,e_{2D,i}),\\
h_{l+1,i} &= \sigma(W_l h_{l,i}+b_l),\quad l=0,\ldots,L-1,\\
r_{\text{high}, i} &= h_{L,i},\\
r_i &= r_{\text{one}, i}+r_{\text{second}, i}+r_{\text{high}, i}.
\end{aligned}
\end{equation}

\subsection{Complexity Analysis}
\textbf{Storage Complexity.}
MOTOR reduces the item-side storage (and loading) complexity from $\mathcal{O}(Nd)$ to $\mathcal{O}(DKd)$, where $D$ is the number of token IDs per item, $K$ is the number of cluster centers per segment, and typically $DK \ll N$. Empirically, MOTOR uses only about 10\%--20\% of the original item-side trainable parameters to learn token representations. Table~\ref{Table:token_number} shows how varying $D$ affects the number of trainable parameters and model performance.

\textbf{Time Complexity.} MOTOR adds a Token Cross Network. Its complexity for first-order, second-order, and higher-order terms is $O(NDd)$, $O(ND^2d)$, and $O(NDd^2)$, respectively, yielding an overall additional complexity of $O(NDd^2)$. Since $d$ and $D$ are typically small and computations are performed in batches, the extra cost is negligible in practice.

\section{Experiments}

\begin{table}[!t]
\centering
% ---- gray caption box (same width as table) ----
\newcommand{\graycap}[1]{%
  \colorbox{gray!15}{\parbox{\dimexpr\linewidth-2\fboxsep\relax}{#1}}%
}

\caption{Statistics of the experimental datasets.}
\label{tab:dataset_stats}

\begin{tabular}{@{}lcccc@{}}
\toprule
\rowcolor{gray!15}
Datasets & \# Users & \# Items & \# Interactions & Sparsity \\
\midrule
Music    & 57202  & 25717 & 184556 & 99.9875\% \\
Baby     & 139722 & 28795 & 501717 & 99.9874\% \\
Office   & 146048 & 38140 & 435675 & 99.9922\% \\
Sports   & 35598  & 18357 & 296337 & 99.9546\% \\
Clothing & 39387  & 23033 & 278677 & 99.9693\% \\
\bottomrule
\end{tabular}
\end{table}

\definecolor{ColorMS}{RGB}{248, 251, 255}
\definecolor{ColorMA}{RGB}{235, 242, 255}
\definecolor{HeadGray}{gray}{0.92}
\definecolor{MaxHL}{RGB}{255, 242, 204}
\setlength{\fboxsep}{0.6pt}
\definecolor{ImprovCol}{RGB}{255, 236, 217}

\begin{table}[ht]
\centering
\caption{Overall performance of MOTOR on the Music dataset.}
\label{table:OverallMusic}
\begin{tabular}{lccccc}
\toprule
\rowcolor{HeadGray}
\textbf{Models} & \textbf{R@10} & \textbf{R@20} & \textbf{N@10} & \textbf{N@20} & \textbf{Improv.} \\
\midrule
\textbf{BPR} & 0.0203 & 0.0310 & 0.0106 & 0.0132 & \cellcolor{ImprovCol}{---} \\
\rowcolor{ColorMS}
\textbf{+MS} & 0.0408 & 0.0602 & 0.0226 & 0.0274 & \cellcolor{ImprovCol}{94.19} \\
\rowcolor{ColorMA}
\textbf{+MA} & 0.0430 & 0.0644 & 0.0241 & 0.0295 & \cellcolor{ImprovCol}{107.74} \\
\midrule
\textbf{LightGCN} & 0.0414 & 0.0588 & 0.0235 & 0.0279 & \cellcolor{ImprovCol}{---} \\
\rowcolor{ColorMS}
\textbf{+MS} & 0.0620 & 0.0897 & 0.0338 & 0.0408 & \cellcolor{ImprovCol}{52.55} \\
\rowcolor{ColorMA}
\textbf{+MA} & 0.0627 & 0.0905 & 0.0345 & 0.0416 & \cellcolor{ImprovCol}{53.91} \\
\midrule
\textbf{VBPR} & 0.0367 & 0.0514 & 0.0206 & 0.0243 & \cellcolor{ImprovCol}{---} \\
\rowcolor{ColorMS}
\textbf{+MS} & 0.0595 & 0.0856 & 0.0323 & 0.0389 & \cellcolor{ImprovCol}{66.54} \\
\rowcolor{ColorMA}
\textbf{+MA} & 0.0589 & 0.0854 & 0.0321 & 0.0388 & \cellcolor{ImprovCol}{66.15} \\
\midrule
\textbf{SLMRec} & 0.0603 & 0.0875 & 0.0319 & 0.0387 & \cellcolor{ImprovCol}{---} \\
\rowcolor{ColorMS}
\textbf{+MS} & 0.0691 & 0.0988 & 0.0373 & 0.0449 & \cellcolor{ImprovCol}{12.91} \\
\rowcolor{ColorMA}
\textbf{+MA} & 0.0688 & 0.0979 & 0.0375 & 0.0449 & \cellcolor{ImprovCol}{11.89} \\
\midrule
\textbf{BM3} & 0.0564 & 0.0783 & 0.0317 & 0.0372 & \cellcolor{ImprovCol}{---} \\
\rowcolor{ColorMS}
\textbf{+MS} & 0.0649 & 0.0988 & 0.0403 & 0.0488 & \cellcolor{ImprovCol}{26.18} \\
\rowcolor{ColorMA}
\textbf{+MA} & 0.0620 & 0.0849 & 0.0383 & 0.0441 & \cellcolor{ImprovCol}{8.43} \\
\midrule
\textbf{FREEDOM} & 0.0715 & 0.1062 & 0.0389 & 0.0477 & \cellcolor{ImprovCol}{---} \\
\rowcolor{ColorMS}
\textbf{+MS} & \textbf{\colorbox{MaxHL}{0.0818}} & \textbf{\colorbox{MaxHL}{0.1164}} & 0.0443 & 0.0531 & \cellcolor{ImprovCol}{9.60} \\
\rowcolor{ColorMA}
\textbf{+MA} & 0.0795 & 0.1157 & 0.0434 & 0.0525 & \cellcolor{ImprovCol}{8.95} \\
\midrule
\textbf{MGCN} & 0.0656 & 0.0926 & 0.0358 & 0.0426 & \cellcolor{ImprovCol}{---} \\
\rowcolor{ColorMS}
\textbf{+MS} & 0.0673 & 0.0939 & 0.0371 & 0.0438 & \cellcolor{ImprovCol}{1.40} \\
\rowcolor{ColorMA}
\textbf{+MA} & 0.0676 & 0.0943 & 0.0376 & 0.0441 & \cellcolor{ImprovCol}{1.62} \\
\midrule
\textbf{LGMRec} & 0.0790 & 0.1131 & 0.0438 & 0.0524 & \cellcolor{ImprovCol}{---} \\
\rowcolor{ColorMS}
\textbf{+MS} & 0.0797 & 0.1154 & 0.0442 & 0.0531 & \cellcolor{ImprovCol}{2.03} \\
\rowcolor{ColorMA}
\textbf{+MA} & 0.0802 & 0.1157 & \textbf{\colorbox{MaxHL}{0.0445}} & \textbf{\colorbox{MaxHL}{0.0534}} & \cellcolor{ImprovCol}{2.30} \\
\midrule
\textbf{GUME} & 0.0733 & 0.1037 & 0.0403 & 0.0480 & \cellcolor{ImprovCol}{---} \\
\rowcolor{ColorMS}
\textbf{+MS} & 0.0775 & 0.1130 & 0.0432 & 0.0517 & \cellcolor{ImprovCol}{8.97} \\
\rowcolor{ColorMA}
\textbf{+MA} & 0.0750 & 0.1110 & 0.0402 & 0.0493 & \cellcolor{ImprovCol}{7.04} \\
\bottomrule
\end{tabular}
\end{table}

\begin{table}[ht]
\centering
\caption{Overall performance of MOTOR on the Baby dataset.}
\label{table:OverallBaby}
\begin{tabular}{lccccc}
\toprule
\rowcolor{HeadGray}
\textbf{Models} & \textbf{R@10} & \textbf{R@20} & \textbf{N@10} & \textbf{N@20} & \textbf{Improv.} \\
\midrule
\textbf{BPR} & 0.0254 & 0.0395 & 0.0133 & 0.0169 & \cellcolor{ImprovCol}{---} \\
\rowcolor{ColorMS}
\textbf{+MS} & 0.0313 & 0.0461 & 0.0172 & 0.0210 & \cellcolor{ImprovCol}{16.71} \\
\rowcolor{ColorMA}
\textbf{+MA} & 0.0319 & 0.0462 & 0.0178 & 0.0214 & \cellcolor{ImprovCol}{16.96} \\
\midrule
\textbf{LightGCN} & 0.0286 & 0.0438 & 0.0155 & 0.0193 & \cellcolor{ImprovCol}{---} \\
\rowcolor{ColorMS}
\textbf{+MS} & 0.0423 & 0.0618 & 0.0235 & 0.0284 & \cellcolor{ImprovCol}{41.10} \\
\rowcolor{ColorMA}
\textbf{+MA} & 0.0423 & 0.0614 & 0.0235 & 0.0283 & \cellcolor{ImprovCol}{40.18} \\
\midrule
\textbf{VBPR} & 0.0323 & 0.0489 & 0.0180 & 0.0222 & \cellcolor{ImprovCol}{---} \\
\rowcolor{ColorMS}
\textbf{+MS} & 0.0386 & 0.0567 & 0.0212 & 0.0257 & \cellcolor{ImprovCol}{15.95} \\
\rowcolor{ColorMA}
\textbf{+MA} & 0.0402 & 0.0578 & 0.0228 & 0.0272 & \cellcolor{ImprovCol}{18.20} \\
\midrule
\textbf{SLMRec} & 0.0490 & 0.0718 & 0.0276 & 0.0333 & \cellcolor{ImprovCol}{---} \\
\rowcolor{ColorMS}
\textbf{+MS} & 0.0547 & 0.0787 & 0.0311 & 0.0372 & \cellcolor{ImprovCol}{9.61} \\
\rowcolor{ColorMA}
\textbf{+MA} & 0.0543 & 0.0782 & 0.0306 & 0.0367 & \cellcolor{ImprovCol}{8.91} \\
\midrule
\textbf{BM3} & 0.0453 & 0.0655 & 0.0256 & 0.0308 & \cellcolor{ImprovCol}{---} \\
\rowcolor{ColorMS}
\textbf{+MS} & 0.0564 & 0.0807 & 0.0313 & 0.0375 & \cellcolor{ImprovCol}{23.21} \\
\rowcolor{ColorMA}
\textbf{+MA} & 0.0542 & 0.0787 & 0.0306 & 0.0368 & \cellcolor{ImprovCol}{20.15} \\
\midrule
\textbf{FREEDOM} & 0.0571 & 0.0813 & 0.0322 & 0.0383 & \cellcolor{ImprovCol}{---} \\
\rowcolor{ColorMS}
\textbf{+MS} & 0.0596 & \textbf{\colorbox{MaxHL}{0.0879}} & \textbf{\colorbox{MaxHL}{0.0347}} & 0.0397 & \cellcolor{ImprovCol}{8.12} \\
\rowcolor{ColorMA}
\textbf{+MA} & 0.0592 & 0.0863 & 0.0341 & 0.0395 & \cellcolor{ImprovCol}{6.15} \\
\midrule
\textbf{MGCN} & 0.0502 & 0.0703 & 0.0291 & 0.0342 & \cellcolor{ImprovCol}{---} \\
\rowcolor{ColorMS}
\textbf{+MS} & 0.0562 & 0.0804 & 0.0318 & 0.0379 & \cellcolor{ImprovCol}{14.37} \\
\rowcolor{ColorMA}
\textbf{+MA} & 0.0567 & 0.0812 & 0.0322 & 0.0380 & \cellcolor{ImprovCol}{15.50} \\
\midrule
\textbf{LGMRec} & 0.0539 & 0.0792 & 0.0300 & 0.0364 & \cellcolor{ImprovCol}{---} \\
\rowcolor{ColorMS}
\textbf{+MS} & 0.0558 & 0.0835 & 0.0311 & 0.0373 & \cellcolor{ImprovCol}{5.43} \\
\rowcolor{ColorMA}
\textbf{+MA} & 0.0564 & 0.0841 & 0.0310 & 0.0375 & \cellcolor{ImprovCol}{6.19} \\
\midrule
\textbf{GUME} & 0.0569 & 0.0805 & 0.0325 & 0.0385 & \cellcolor{ImprovCol}{---} \\
\rowcolor{ColorMS}
\textbf{+MS} & 0.0595 & 0.0842 & 0.0336 & 0.0399 & \cellcolor{ImprovCol}{4.60} \\
\rowcolor{ColorMA}
\textbf{+MA} & \textbf{\colorbox{MaxHL}{0.0601}} & 0.0857 & 0.0339 & \textbf{\colorbox{MaxHL}{0.0404}} & \cellcolor{ImprovCol}{6.46} \\
\bottomrule
\end{tabular}
\end{table}

\begin{table}[ht]
\centering
\caption{Overall performance of MOTOR on the Office dataset.}
\label{table:OverallOffice}
\begin{tabular}{lccccc}
\toprule
\rowcolor{HeadGray}
\textbf{Models} & \textbf{R@10} & \textbf{R@20} & \textbf{N@10} & \textbf{N@20} & \textbf{Improv.} \\
\midrule
\textbf{BPR} & 0.0178 & 0.0250 & 0.0108 & 0.0126 & \cellcolor{ImprovCol}{---} \\
\rowcolor{ColorMS}
\textbf{+MS} & 0.0406 & 0.0561 & 0.0238 & 0.0277 & \cellcolor{ImprovCol}{124.40} \\
\rowcolor{ColorMA}
\textbf{+MA} & 0.0424 & 0.0578 & 0.0247 & 0.0286 & \cellcolor{ImprovCol}{131.20} \\
\midrule
\textbf{LightGCN} & 0.0333 & 0.0434 & 0.0213 & 0.0239 & \cellcolor{ImprovCol}{---} \\
\rowcolor{ColorMS}
\textbf{+MS} & 0.0561 & 0.0774 & 0.0331 & 0.0385 & \cellcolor{ImprovCol}{78.34} \\
\rowcolor{ColorMA}
\textbf{+MA} & 0.0567 & 0.0774 & 0.0331 & 0.0383 & \cellcolor{ImprovCol}{78.34} \\
\midrule
\textbf{VBPR} & 0.0388 & 0.0494 & 0.0254 & 0.0281 & \cellcolor{ImprovCol}{---} \\
\rowcolor{ColorMS}
\textbf{+MS} & 0.0578 & 0.0755 & 0.0349 & 0.0394 & \cellcolor{ImprovCol}{52.83} \\
\rowcolor{ColorMA}
\textbf{+MA} & 0.0554 & 0.0731 & 0.0335 & 0.0380 & \cellcolor{ImprovCol}{47.98} \\
\midrule
\textbf{SLMRec} & 0.0557 & 0.0721 & 0.0346 & 0.0387 & \cellcolor{ImprovCol}{---} \\
\rowcolor{ColorMS}
\textbf{+MS} & 0.0664 & 0.0899 & 0.0386 & 0.0446 & \cellcolor{ImprovCol}{24.69} \\
\rowcolor{ColorMA}
\textbf{+MA} & 0.0650 & 0.0886 & 0.0378 & 0.0438 & \cellcolor{ImprovCol}{22.88} \\
\midrule
\textbf{BM3} & 0.0501 & 0.0658 & 0.0313 & 0.0353 & \cellcolor{ImprovCol}{---} \\
\rowcolor{ColorMS}
\textbf{+MS} & 0.0574 & 0.0837 & 0.0315 & 0.0382 & \cellcolor{ImprovCol}{27.20} \\
\rowcolor{ColorMA}
\textbf{+MA} & 0.0625 & 0.0882 & 0.0361 & 0.0426 & \cellcolor{ImprovCol}{34.04} \\
\midrule
\textbf{FREEDOM} & 0.0721 & 0.0973 & 0.0421 & 0.0485 & \cellcolor{ImprovCol}{---} \\
\rowcolor{ColorMS}
\textbf{+MS} & 0.0765 & 0.1063 & 0.0427 & 0.0502 & \cellcolor{ImprovCol}{9.25} \\
\rowcolor{ColorMA}
\textbf{+MA} & 0.0743 & 0.1036 & 0.0425 & 0.0491 & \cellcolor{ImprovCol}{6.47} \\
\midrule
\textbf{MGCN} & 0.0714 & 0.0928 & 0.0427 & 0.0481 & \cellcolor{ImprovCol}{---} \\
\rowcolor{ColorMS}
\textbf{+MS} & 0.0775 & 0.1057 & 0.0443 & 0.0514 & \cellcolor{ImprovCol}{13.90} \\
\rowcolor{ColorMA}
\textbf{+MA} & 0.0789 & 0.1074 & 0.0458 & 0.0530 & \cellcolor{ImprovCol}{15.73} \\
\midrule
\textbf{LGMRec} & 0.0675 & 0.0951 & 0.0374 & 0.0444 & \cellcolor{ImprovCol}{---} \\
\rowcolor{ColorMS}
\textbf{+MS} & 0.0686 & 0.0998 & 0.0397 & 0.0453 & \cellcolor{ImprovCol}{4.94} \\
\rowcolor{ColorMA}
\textbf{+MA} & 0.0682 & 0.0994 & 0.0389 & 0.0449 & \cellcolor{ImprovCol}{4.52} \\
\midrule
\textbf{GUME} & 0.0662 & 0.0890 & 0.0386 & 0.0444 & \cellcolor{ImprovCol}{---} \\
\rowcolor{ColorMS}
\textbf{+MS} & \textbf{\colorbox{MaxHL}{0.0800}} & \textbf{\colorbox{MaxHL}{0.1078}} & \textbf{\colorbox{MaxHL}{0.0468}} & \textbf{\colorbox{MaxHL}{0.0538}} & \cellcolor{ImprovCol}{21.12} \\
\rowcolor{ColorMA}
\textbf{+MA} & 0.0785 & 0.1043 & 0.0467 & 0.0532 & \cellcolor{ImprovCol}{17.19} \\
\bottomrule
\end{tabular}
\end{table}

\begin{table}[ht]
\centering
\caption{The cold-start recommendation results on Sports and Clothing.}
\label{Table:Cold-Start}
\begin{tabular}{lcccc}
\toprule
\rowcolor{gray!15}
\textbf{Models} & \multicolumn{2}{c}{\textbf{Sports}} & \multicolumn{2}{c}{\textbf{Clothing}} \\
\rowcolor{gray!15}
 & \textbf{R@20} & \textbf{N@20} & \textbf{R@20} & \textbf{N@20} \\
\midrule

BPR      & 0.0061 & 0.0024 & 0.0044 & 0.0017 \\
\rowcolor{blue!10}
+MS      & 0.0257 & 0.0110 & 0.0353 & 0.0150 \\
\midrule

LightGCN & 0.0014 & 0.0012 & 0.0029 & 0.0012 \\
\rowcolor{blue!10}
+MS      & 0.0135 & 0.0055 & 0.0276 & 0.0108 \\
\midrule

VBPR     & 0.0098 & 0.0039 & 0.0167 & 0.0071 \\
\rowcolor{blue!10}
+MS      & 0.0243 & 0.0104 & \textbf{0.0375} & 0.0163 \\
\midrule

BM3      & 0.0032 & 0.0012 & 0.0041 & 0.0013 \\
\rowcolor{blue!10}
+MS      & 0.0113 & 0.0047 & 0.0141 & 0.0062 \\
\midrule

FREEDOM  & 0.0099 & 0.0033 & 0.0213 & 0.0046 \\
\rowcolor{blue!10}
+MS      & 0.0191 & 0.0129 & 0.0327 & 0.0147 \\
\midrule

MGCN     & 0.0072 & 0.0023 & 0.0068 & 0.0022 \\
\rowcolor{blue!10}
+MS      & 0.0206 & 0.0094 & 0.0248 & 0.0129 \\
\midrule

GUME     & 0.0188 & 0.0063 & 0.0226 & 0.0085 \\
\rowcolor{blue!10}
+MS      & \textbf{0.0262} & \textbf{0.0115} & 0.0361 & \textbf{0.0166} \\
\bottomrule
\end{tabular}
\end{table}

We conduct comprehensive experiments to evaluate the performance of MOTOR and answer the following questions:
\begin{itemize}
    \item \textbf{RQ1}: What performance gains do ID-free recommendation models enhanced by MOTOR achieve compared to their ID-based versions in the common recommendation scenario?
    \item \textbf{RQ2}: How do the ID-free recommender and original ID-based models perform differently for items with varying numbers of interactions (e.g., cold-start items, popular items)?
    \item \textbf{RQ3}: How do the token numbers impact the number of trainable parameters and model performance?
    \item \textbf{RQ4}: How do the different modules influence the performance of MOTOR?
    % \item \textbf{RQ4}: How does MOTOR perform about spatial optimization compared with original models under different token number settings?
\end{itemize}

\subsection{Experimental setting}
\subsubsection{\textbf{Datasets}}
We evaluate on five Amazon datasets~\cite{ni2019justifying}: Music, Baby, Office, Sports and Clothing. 
Following MMRec~\cite{MMrec}, we use 4096-d CNN visual features and 384-d Sentence-Transformer text features, with an 8/1/1 train/val/test split per user.
To thoroughly evaluate the performance of MOTOR, we design two types of scenarios: 
\begin{itemize}
    \item \textbf{Common Recommendation (Music, Baby, Office)}: we apply 1-core filtering to ensure each user and item has at least one interaction. In this scenario, both the training and test sets contain items with diverse interaction records, including long-tail items and popular items;
    \item \textbf{Cold-start Recommendation (Sports, Clothing)}: The items in the test sets are cold-start items without interaction records. In this scenario, we evaluate the model's recommendation performance for cold-start items.
\end{itemize}

\subsubsection{\textbf{BackBone Models}}
We integrate MOTOR into nine ID-based recommenders. The general recommenders are \textbf{BPR}~\cite{rendle2012bpr} and \textbf{LightGCN}~\cite{he2020lightgcn}, which do not utilize multimodal features as supervision signals during training. We also include seven multimodal recommendation models, namely \textbf{VBPR}~\cite{vbpr}, \textbf{SLMRec}~\cite{SLMRec}, \textbf{BM3}~\cite{BM3}, \textbf{FREEDOM}~\cite{Freedom}, \textbf{MGCN}~\cite{MGCN}, \textbf{LGMRec}~\cite{guo2023lgmrec}, and \textbf{GLUE}~\cite{GUME}, which incorporate multimodal features to learn user and item representations.
Especially, for General Recommendation models, the final user and item representation can be denoted as $h_u,h_i=f_{\theta}(e_u,e_i)$, the MOTOR-enhanced model is denoted as $h_u,h_i=f_{\theta}(e_u,g_{\phi}(t_i^t,t_i^v))$.

\subsubsection{\textbf{Evaluation and Implementation Details}}
We evaluate top-$K$ recommendation using Recall@$K$ (R@$K$) and NDCG@$K$ (N@$K$), reporting results for $K\in\{10,20\}$. We implement Optimized Product Quantization with Faiss ANNS~\cite{Faiss} and integrate MOTOR into ID-based recommenders via MMRec~\cite{MMrec}. Vector quantization is performed once and reused across downstream models. Following prior work~\cite{BM3,Freedom,MGCN}, we set the embedding dimension of users, items, and tokens to 64, initialize embeddings with Xavier~\cite{Xavier}, and optimize with Adam~\cite{kingma2014adam} using a learning rate of 0.001. For fair comparison, we tune each baseline according to its original paper. MOTOR adds only one hyperparameter, the token number searched over $\{2,4,8,16\}$, while fixing the number of cluster centers to 256.

% \usepackage{multirow}
% \usepackage{multirow}
% \usepackage{multirow}
% \usepackage{multirow}

\begin{figure}[ht]
  \centering
  \includegraphics[width=\linewidth]{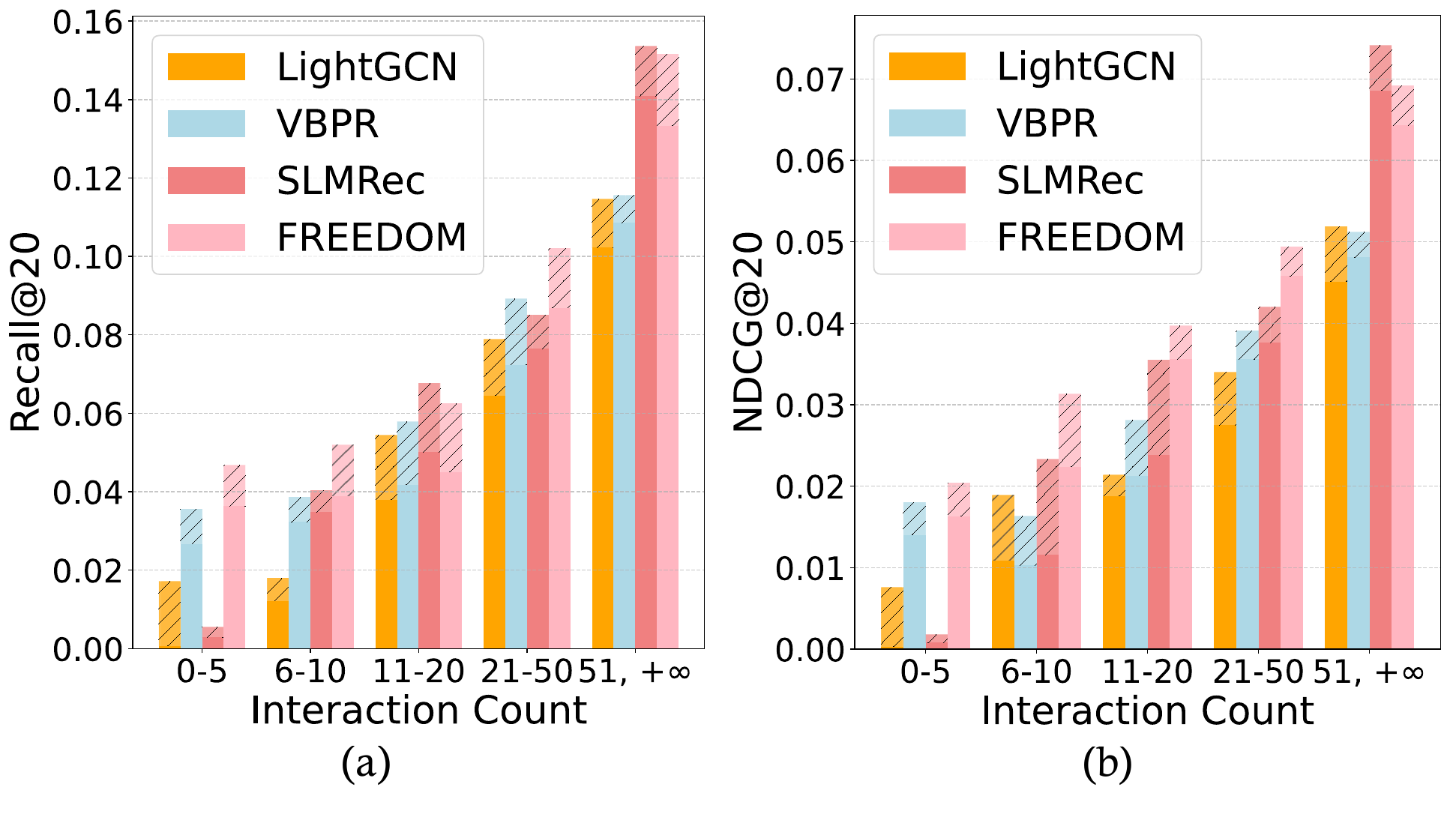}
  \caption{
 The Recall@20 (a) and NDCG@20 (b) of four models for items with diverse interaction counts. The shaded areas represent the performance improvement.
  }
  \label{fig:Item_recall20_baby}
% \vspace{-5pt}
\end{figure}

\subsection{Overall Performance Comparison (RQ1)}
We conduct common recommendations on the Music, Baby, and Office datasets and report results for multiple backbones in Tables~\ref{table:OverallMusic}--\ref{table:OverallOffice}. Three key findings emerge. 

\begin{itemize}
\item \textbf{Consistent gains across backbones.} MOTOR yields substantial improvements regardless of the backbone. The magnitude of the gain depends on the underlying model: it markedly improves weaker backbones (e.g., BPR, LightGCN, VBPR) and further enhances strong baselines, achieving state-of-the-art performance with advanced models (e.g., FREEDOM, GUME).  

\item \textbf{Largest improvements for general recommenders.} The benefits are most pronounced for general recommenders; MOTOR-BPR more than doubles Recall@20 on Music and Office. Notably, MOTOR does not directly inject multimodal features into these models. Instead, by tokenizing multimodal signals and converting ID-based recommendation into an ID-free formulation, it alleviates issues such as information isolation and cold start while implicitly leveraging semantic information encoded by token IDs.  

\item \textbf{MA vs.\ MS Token Cross Networks.} We compare two Token Cross Network variants: modal-agnostic (MA) and modal-specific (MS). Both effectively model interactions among multimodal tokens, and their relative advantage depends on the backbone and data distribution. For instance, MA consistently outperforms MS for BPR and MGCN, whereas MS is superior for SLMRec and FREEDOM; for other backbones, their performance is comparable.  
\end{itemize}

\subsection{Recommendation Performance for Cold-start and Popular Items}
We evaluate MOTOR on cold-start and popularity-stratified items. On Sports and Clothing (Table~\ref{Table:Cold-Start}), all backbones deteriorate on cold-start items with no interactions, whereas MOTOR markedly improves performance by sharing multimodal tokens to enrich cold-item representations.

On Baby, we bucket test items by training interaction counts ($[0\!-\!5]$, $[6\!-\!10]$, $[11\!-\!20]$, $[20\!-\!50]$, $[50,\infty)$). Figure~\ref{fig:Item_recall20_baby} shows consistent Recall@20 gains across all buckets, including both long-tail and popular items; even for $>50$ interactions, MOTOR still improves over ID-based models.

\begin{table}[ht]
\centering
\caption{The trainable parameters and Recall@20 of original and MOTOR-enhanced models under different uni-modal token numbers in the Music dataset. The second row shows Item Embedding parameter variations. \textcolor{gain}{Orange} and \textcolor{loss}{cyan} indicate performance gains or reduced parameters compared to the corresponding backbone, with deeper shades meaning greater relative changes.}
\label{Table:token_number}

\begin{tabular}{c c ccccc}
\toprule
\rowcolor{gray!15}
 &  & \multicolumn{1}{c}{Original} &
 \multicolumn{1}{c}{2 Tokens} &
 \multicolumn{1}{c}{4 Tokens} &
 \multicolumn{1}{c}{8 Tokens} &
 \multicolumn{1}{c}{16 Tokens} \\
\midrule

\rowcolor{gray!15}
\multicolumn{1}{c}{All Models} &
\multicolumn{1}{c}{\makecell{Item Embedding\\ Parameters}} &
1,645,888 &
\cellcolor{loss!90}65,536 &
\cellcolor{loss!80}131,072 &
\cellcolor{loss!70}262,144 &
\cellcolor{loss!60}524,288 \\
\midrule\midrule

\multirow{2}{*}{LightGCN} &
\makecell{Trainable\\ Params} &
5,306,816 &
\cellcolor{loss!50}3,743,236 &
\cellcolor{loss!31}3,825,352 &
\cellcolor{loss!19}3,989,776 &
\cellcolor{loss!10}4,318,496 \\
\cmidrule(lr){2-7}
& Recall@20 &
0.0588 &
\cellcolor{gain!30}0.0782 &
\cellcolor{gain!50}0.0897 &
\cellcolor{gain!45}0.0876 &
\cellcolor{gain!33}0.0787 \\
\midrule

\multirow{2}{*}{VBPR} &
\makecell{Trainable\\ Params} &
9,254,528 &
\cellcolor{loss!50}7,502,532 &
\cellcolor{loss!35}7,773,128 &
\cellcolor{loss!27}7,937,488 &
\cellcolor{loss!13}8,266,208 \\
\cmidrule(lr){2-7}
& Recall@20 &
0.0514 &
\cellcolor{gain!20}0.0663 &
\cellcolor{gain!30}0.0767 &
\cellcolor{gain!40}0.0856 &
\cellcolor{gain!35}0.0812 \\
\midrule

\multirow{2}{*}{SLMRec} &
\makecell{Trainable\\ Params} &
5,639,168 &
\cellcolor{loss!50}4,075,588 &
\cellcolor{loss!30}4,157,768 &
\cellcolor{loss!25}4,322,128 &
\cellcolor{loss!15}4,650,848 \\
\cmidrule(lr){2-7}
& Recall@20 &
0.0875 &
\cellcolor{gain!5}0.0897 &
\cellcolor{gain!15}0.0985 &
\cellcolor{gain!20}0.0988 &
\cellcolor{gain!10}0.0962 \\
\bottomrule
\end{tabular}
\end{table}

\begin{table}[ht]
\centering
\caption{The Ablation Study about modal tokens and Token Cross Network. We adopt the MS TCN structure.}
\label{Table:AblationStudy}

\begingroup
\setlength{\tabcolsep}{4pt}

\begin{tabular}{c p{3.35cm} cc cc}
\toprule
\rowcolor{gray!15}
\textbf{Models} & \textbf{Settings} & \multicolumn{2}{c}{\textbf{Music}} & \multicolumn{2}{c}{\textbf{Office}} \\
\rowcolor{gray!15}
 &  & \textbf{R@20} & \textbf{N@20} & \textbf{R@20} & \textbf{N@20} \\
\midrule
\cmidrule(lr){3-4}\cmidrule(lr){5-6}

\multirow{8}{*}{VBPR}
& Original         & 0.0514 & 0.0243 & 0.0494 & 0.0281 \\
& w/o Text Tokens  & 0.0655 & 0.0302 & 0.0560 & 0.0289 \\
& w/o Image Tokens & 0.0817 & 0.0370 & 0.0729 & 0.0344 \\
& w/o one-order    & 0.0697 & 0.0334 & 0.0629 & 0.0315 \\
& w/o second-order & 0.0828 & 0.0375 & 0.0733 & 0.0363 \\
& w/o high-order   & 0.0837 & 0.0381 & 0.0742 & 0.0379 \\
& w/o TCN          & 0.0791 & 0.0344 & 0.0682 & 0.0331 \\
\rowcolor{blue!10}
& MOTOR            & \textbf{0.0856} & \textbf{0.0389} & \textbf{0.0755} & \textbf{0.0394} \\
\midrule\midrule

\multirow{8}{*}{SLMRec}
& Original         & 0.0875 & 0.0387 & 0.0721 & 0.0387 \\
& w/o Text Tokens  & 0.0898 & 0.0396 & 0.0783 & 0.0399 \\
& w/o Image Tokens & 0.0963 & 0.0438 & 0.0848 & 0.0405 \\
& w/o one-order    & 0.0866 & 0.0374 & 0.0715 & 0.0379 \\
& w/o second-order & 0.0944 & 0.0425 & 0.0837 & 0.0419 \\
& w/o high-order   & 0.0939 & 0.0417 & 0.0821 & 0.0415 \\
& w/o TCN          & 0.0905 & 0.0407 & 0.0812 & 0.0407 \\
\rowcolor{blue!10}
& MOTOR            & \textbf{0.0988} & \textbf{0.0449} & \textbf{0.0899} & \textbf{0.0446} \\
\midrule\midrule

\multirow{8}{*}{FREEDOM}
& Original         & 0.1062 & 0.0477 & 0.0973 & 0.0485 \\
& w/o Text Tokens  & 0.1102 & 0.0484 & 0.0995 & 0.0492 \\
& w/o Image Tokens & 0.1139 & 0.0510 & 0.1014 & 0.0497 \\
& w/o one-order    & 0.0987 & 0.0454 & 0.0952 & 0.0469 \\
& w/o second-order & 0.1133 & 0.0509 & 0.1005 & 0.0485 \\
& w/o high-order   & 0.1152 & 0.0522 & 0.1039 & 0.0491 \\
& w/o TCN          & 0.1098 & 0.0481 & 0.0988 & 0.0489 \\
\rowcolor{blue!10}
& MOTOR            & \textbf{0.1164} & \textbf{0.0531} & \textbf{0.1063} & \textbf{0.0502} \\
\bottomrule
\end{tabular}
\endgroup
\end{table}

\subsection{Trainable Paremeters and Performance Under Different Token Numbers (RQ3)}\label{sec:RQ4}
MOTOR substitutes ID embeddings with learnable Token Embedding tables. Table \ref{Table:token_number} shows trainable parameters and performance under various token settings. ID-free models typically outperform ID-based counterparts while reducing parameters significantly. With 4 uni-modal tokens, ID-free LightGCN, VBPR, and SLMRec achieve parameter reductions of 27.9\%, 16.0\%, and 26.3\% respectively, with performance gains of 52.6\%, 49.2\%, and 12.6\%.

Optimal performance occurs with 4 or 8 unimodal tokens. Fewer tokens constrain embedding space and increase collision probability, while excessive tokens enhance expressiveness but introduce modality-related noise, causing unrelated items to negatively influence each other through shared token IDs. This highlights the critical balance between model expressiveness and noise management in token assignment.

\subsection{Ablation Study (RQ4)}\label{sec:ablation}
We analyze how each of the proposed components affects the performance of MOTOR using three backbone models (e.g., VBPR~\cite{vbpr}, SLMRec~\cite{SLMRec}, and FREEDOM~\cite{Freedom}), and the experimental results are shown in Table \ref{Table:AblationStudy}. We examine multimodal tokens by removing specific modality tokens and evaluate the TCN by separately removing one-order, second-order, and high-order components, as well as replacing TCN with Mean operation. Key findings include: \textbf{(i)} Removing either image or text tokens degrades performance, with text tokens being more crucial as `w/o Image Tokens' achieves better results. \textbf{(ii)} One-order representation is most critical in TCN, with `w/o one-order' causing the largest performance drop, while second-order and high-order components enhance model expressiveness. \textbf{(iii)} Replacing TCN with Mean operation (`w/o TCN') still maintains superior performance, confirming that token embeddings capture essential semantic information.

\section{Conclusion}
We identify three key limitations of existing ID-based multimodal recommenders: information isolation, cold-start, and storage overhead. To address these challenges, we propose MOTOR, which represents items via learnable multimodal tokens, thereby converting ID-based recommenders into ID-free systems. A token-sharing mechanism facilitates knowledge transfer among related items, enhances representations for cold-start items, and removes the need for a large item ID embedding table. Moreover, by treating tokens as implicit item features, we introduce a Token Cross Network (TCN) to efficiently capture first-, second-, and higher-order interactions among multimodal tokens, yielding more expressive representations without substantial computational or storage costs. Extensive experiments on nine representative recommendation models and five public datasets show that MOTOR consistently achieves significant performance gains.

\begin{credits}
\subsubsection{\ackname}
This paper is supported by National Natural Science Foundation
of China (624B2096, 72595872, 72542012, 62322603).

\subsubsection{\discintname}
The authors have no competing interests to declare that are relevant to the content of this article.
\end{credits}

\bibliographystyle{splncs04}
\bibliography{myref}

\begin{thebibliography}{10}
\providecommand{\url}[1]{\texttt{#1}}
\providecommand{\urlprefix}{URL }
\providecommand{\doi}[1]{https://doi.org/#1}

\bibitem{cheng2016widedeeplearning}
Cheng, H.T., Koc, L., Harmsen, J., Shaked, T., Chandra, T., Aradhye, H., Anderson, G., Corrado, G., Chai, W., Ispir, M., Anil, R., Haque, Z., Hong, L., Jain, V., Liu, X., Shah, H.: Wide \& deep learning for recommender systems (2016)

\bibitem{vit}
Dosovitskiy, A., Beyer, L., Kolesnikov, A., Weissenborn, D., Zhai, X., Unterthiner, T., Dehghani, M., Minderer, M., Heigold, G., Gelly, S., et~al.: An image is worth 16x16 words: Transformers for image recognition at scale. arXiv preprint arXiv:2010.11929  (2020)

\bibitem{OPQ}
Ge, T., He, K., Ke, Q., Sun, J.: Optimized product quantization. IEEE transactions on pattern analysis and machine intelligence  \textbf{36}(4),  744--755 (2013)

\bibitem{Xavier}
Glorot, X., Bengio, Y.: Understanding the difficulty of training deep feedforward neural networks. In: Proceedings of the Thirteenth International Conference on Artificial Intelligence and Statistics. pp. 249--256. JMLR Workshop and Conference Proceedings (2010)

\bibitem{DeepFM}
Guo, H., Tang, R., Ye, Y., Li, Z., He, X.: Deepfm: A factorization-machine based neural network for ctr prediction (2017)

\bibitem{guo2023lgmrec}
Guo, Z., Li, J., Li, G., Wang, C., Shi, S., Ruan, B.: Lgmrec: Local and global graph learning for multimodal recommendation (2023)

\bibitem{resnet}
He, K., Zhang, X., Ren, S., Sun, J.: Deep residual learning for image recognition. In: Proceedings of the IEEE conference on computer vision and pattern recognition. pp. 770--778 (2016)

\bibitem{vbpr}
He, R., McAuley, J.: Vbpr: visual bayesian personalized ranking from implicit feedback. In: Proceedings of the AAAI conference on artificial intelligence (2016)

\bibitem{he2020lightgcn}
He, X., Deng, K., Wang, X., Li, Y., Zhang, Y., Wang, M.: Lightgcn: Simplifying and powering graph convolution network for recommendation. In: Proceedings of the 43rd International ACM SIGIR Conference on Research and Development in Information Retrieval. pp. 639--648 (2020)

\bibitem{UniSRec}
Hou, Y., Mu, S., Zhao, W.X., Li, Y., Ding, B., Wen, J.R.: Towards universal sequence representation learning for recommender systems (2022)

\bibitem{Faiss}
Johnson, J., Douze, M., Jégou, H.: Billion-scale similarity search with gpus (2017)

\bibitem{kingma2014adam}
Kingma, D.P., Ba, J.: Adam: A method for stochastic optimization. arXiv preprint arXiv:1412.6980  (2014)

\bibitem{Recformer}
Li, J., Wang, M., Li, J., Fu, J., Shen, X., Shang, J., McAuley, J.: Text is all you need: Learning language representations for sequential recommendation. In: Proceedings of the 29th ACM SIGKDD Conference on Knowledge Discovery and Data Mining. p. 1258–1267. KDD '23, Association for Computing Machinery, New York, NY, USA (2023)

\bibitem{GUME}
Lin, G., Meng, Z., Wang, D., Long, Q., Zhou, Y., Xiao, M.: Gume: Graphs and user modalities enhancement for long-tail multimodal recommendation (2024)

\bibitem{liu2017deepstyle}
Liu, Q., Wu, S., Wang, L.: Deepstyle: Learning user preferences for visual recommendation. In: Proceedings of the 40th international acm sigir conference on research and development in information retrieval. pp. 841--844 (2017)

\bibitem{AlignRec}
Liu, Y., Zhang, K., Ren, X., Huang, Y., Jin, J., Qin, Y., Su, R., Xu, R., Yu, Y., Zhang, W.: Alignrec: Aligning and training in multimodal recommendations. In: Proceedings of the 33rd ACM International Conference on Information and Knowledge Management. pp. 1503--1512 (2024)

\bibitem{ni2019justifying}
Ni, J., Li, J., McAuley, J.: Justifying recommendations using distantly-labeled reviews and fine-grained aspects. In: Proceedings of the 2019 Conference on Empirical Methods in Natural Language Processing and the 9th International Joint Conference on Natural Language Processing. pp. 188--197 (2019)

\bibitem{CLIP}
Radford, A., Kim, J.W., Hallacy, C., Ramesh, A., Goh, G., Agarwal, S., Sastry, G., Askell, A., Mishkin, P., Clark, J., et~al.: Learning transferable visual models from natural language supervision. In: International conference on machine learning. pp. 8748--8763. PMLR (2021)

\bibitem{GPT2}
Radford, A., Wu, J., Child, R., Luan, D., Amodei, D., Sutskever, I., et~al.: Language models are unsupervised multitask learners. OpenAI blog  \textbf{1}(8), ~9 (2019)

\bibitem{T5}
Raffel, C., Shazeer, N., Roberts, A., Lee, K., Narang, S., Matena, M., Zhou, Y., Li, W., Liu, P.J.: Exploring the limits of transfer learning with a unified text-to-text transformer (2023)

\bibitem{sentencebert}
Reimers, N., Gurevych, I.: Sentence-bert: Sentence embeddings using siamese bert-networks. arXiv preprint arXiv:1908.10084  (2019)

\bibitem{rendle2012bpr}
Rendle, S., Freudenthaler, C., Gantner, Z., Schmidt-Thieme, L.: Bpr: Bayesian personalized ranking from implicit feedback. arXiv preprint arXiv:1205.2618  (2012)

\bibitem{AutoGraph}
Shan, R., Lin, J., Zhu, C., Chen, B., Zhu, M., Zhang, K., Zhu, J., Tang, R., Yu, Y., Zhang, W.: An automatic graph construction framework based on large language models for recommendation. In: Proceedings of the 31st ACM SIGKDD Conference on Knowledge Discovery and Data Mining V.2. p. 4806–4817. KDD '25, Association for Computing Machinery, New York, NY, USA (2025). \doi{10.1145/3711896.3737192}, \url{https://doi.org/10.1145/3711896.3737192}

\bibitem{VGG}
Simonyan, K., Zisserman, A.: Very deep convolutional networks for large-scale image recognition. arXiv preprint arXiv:1409.1556  (2014)

\bibitem{SLMRec}
Tao, Z., Liu, X., Xia, Y., Wang, X., Yang, L., Huang, X., Chua, T.S.: Self-supervised learning for multimedia recommendation. IEEE Transactions on Multimedia  (2022)

\bibitem{bert}
Vaswani, A., Shazeer, N., Parmar, N., Uszkoreit, J., Jones, L., Gomez, A.N., Kaiser, {\L}., Polosukhin, I.: Attention is all you need. Advances in neural information processing systems  \textbf{30} (2017)

\bibitem{dualgnn}
Wang, Q., Wei, Y., Yin, J., Wu, J., Song, X., Nie, L.: Dualgnn: Dual graph neural network for multimedia recommendation. IEEE Transactions on Multimedia  (2021)

\bibitem{DCN}
Wang, R., Fu, B., Fu, G., Wang, M.: Deep \& cross network for ad click predictions (2017)

\bibitem{GRCN}
Wei, Y., Wang, X., Nie, L., He, X., Chua, T.S.: Graph-refined convolutional network for multimedia recommendation with implicit feedback. In: Proceedings of the 28th ACM international conference on multimedia. pp. 3541--3549 (2020)

\bibitem{wei2019mmgcn}
Wei, Y., Wang, X., Nie, L., He, X., Hong, R., Chua, T.S.: Mmgcn: Multi-modal graph convolution network for personalized recommendation of micro-video. In: Proceedings of the 27th ACM international conference on multimedia (2019)

\bibitem{wu2021towards}
Wu, Q., Zhang, H., Gao, X., Yan, J., Zha, H.: Towards open-world recommendation: An inductive model-based collaborative filtering approach. In: International Conference on Machine Learning. pp. 11329--11339. PMLR (2021)

\bibitem{INMO}
Wu, Y., Cao, Q., Shen, H., Tao, S., Cheng, X.: Inmo: A model-agnostic and scalable module for inductive collaborative filtering. In: Proceedings of the 45th International ACM SIGIR Conference on Research and Development in Information Retrieval. pp. 91--101. SIGIR '22, Association for Computing Machinery, New York, NY, USA (2022)

\bibitem{MGCN}
Yu, P., Tan, Z., Lu, G., Bao, B.K.: Multi-view graph convolutional network for multimedia recommendation. arXiv preprint arXiv:2308.03588  (2023)

\bibitem{IDvsMoRec}
Yuan, Z., Yuan, F., Song, Y., Li, Y., Fu, J., Yang, F., Pan, Y., Ni, Y.: Where to go next for recommender systems? id- vs. modality-based recommender models revisited (2023)

\bibitem{DREAM}
Zhang, K., Qin, Y., Su, R., Liu, Y., Jin, J., Zhang, W., Yu, Y.: {DREAM}: A dual representation learning framework for multimodal recommendation (2024)

\bibitem{Dual-Aligned}
Zhang, K., Qin, Y., Su, R., Liu, Y., Zhang, W., Yu, Y.: A dual-aligned model for multimodal recommendation. In: He, X., Ren, Z., Tang, R. (eds.) Information Retrieval. pp. 14--27. Springer Nature Singapore, Singapore (2025)

\bibitem{MMrec}
Zhou, H., Zhou, X., Zeng, Z., Zhang, L., Shen, Z.: A comprehensive survey on multimodal recommender systems: Taxonomy, evaluation, and future directions. arXiv preprint arXiv:2302.04473  (2023)

\bibitem{Freedom}
Zhou, X., Shen, Z.: A tale of two graphs: Freezing and denoising graph structures for multimodal recommendation. In: Proceedings of the 31st ACM International Conference on Multimedia. MM ’23, ACM (Oct 2023)

\bibitem{BM3}
Zhou, X., Zhou, H., Liu, Y., Zeng, Z., Miao, C., Wang, P., You, Y., Jiang, F.: Bootstrap latent representations for multi-modal recommendation. In: Proceedings of the ACM Web Conference 2023. pp. 845--854 (2023)

\end{thebibliography}

\end{document}